\documentclass[conference]{IEEEtran}
\usepackage[utf8]{inputenc}
\IEEEoverridecommandlockouts
\usepackage{listings}
\usepackage{cite}
\usepackage{amsmath, amssymb, amsfonts, amsthm, mathrsfs}
\usepackage{xcolor}
\usepackage{textcomp}
\usepackage{booktabs}
\usepackage{graphicx}
\usepackage{tabularx}
\usepackage{array}
\usepackage{url}
\usepackage{longtable}

\usepackage{caption}
\usepackage{tikz}
\usepackage{hyperref}
\usepackage{pdflscape} % For landscape tables
\usetikzlibrary{mindmap}
\usetikzlibrary{shapes.geometric, arrows, positioning}
\usepackage{soul}
\usepackage{enumitem}
\usepackage[colorinlistoftodos]{todonotes}
\usepackage{subcaption}

\hypersetup{
    colorlinks=true,
    linkcolor=blue,
    citecolor=blue,
    urlcolor=blue,
}

\title{Audio Foundation Models for Education:\\ 
New Opportunities and the Road Ahead}
\title{Signal Processing Meets Generative AI:\\ Using Audio Foundation Models in Education}
\title{Teaching Signal Processing in the AI Era: The Role of Audio Foundation Models}
\title{Generative AI in Signal Processing Education:\\ An Audio Foundation Model Based Approach}

\author{
 Muhammad Salman Khan\textsuperscript{1}, 
 Ahmad Ullah\textsuperscript{2}, 
 Siddique Latif\textsuperscript{3}, 
 Junaid Qadir\textsuperscript{2} \\
 \textsuperscript{1}Department of Electrical Engineering, Qatar University,\\
  \textsuperscript{2}Department of Computer Science and Engineering, Qatar University,\\
  \textsuperscript{3}Queensland University of Technology (QUT), Australia,\\
 salman@qu.edu.qa, au2315111@student.qu.edu.qa, siddique.latif@usq.edu.au,  jqadir@qu.edu.qa
}
\begin{document}
\maketitle
\begin{abstract}
Audio Foundation Models (AFMs), a specialized category of Generative AI (GenAI), have the potential to transform signal processing (SP) education by integrating core applications such as speech and audio enhancement, denoising, source separation, feature extraction, automatic classification, and real-time signal analysis into learning and research. This paper introduces \textit{SPEduAFM}, a conceptual AFM tailored for SP education that fuses traditional SP principles with GenAI-driven innovations. Through an envisioned case study, we outline how AFMs can enable a range of applications, including automated lecture transcription, interactive demonstrations, and inclusive learning tools, showcasing their potential to transform abstract concepts into engaging practical experiences. This paper also addresses challenges such as ethics, explainability, and customization by highlighting dynamic, real-time auditory interactions that foster experiential and authentic learning. By presenting SPEduAFM as a forward-looking vision, our goal is to inspire broader adoption of GenAI in engineering education, enhancing accessibility, engagement, and innovation in the classroom and beyond.
\end{abstract}

\section{Introduction}

Generative Artificial Intelligence (GenAI) has emerged as a transformative technology in multiple domains, demonstrating its potential to redefine educational paradigms. In the field of audio signal processing, the advent of Audio Foundation Models (AFMs), a subset of GenAI, represents a significant leap forward. These models enable advanced capabilities such as audio synthesis, real-time analysis, and interpretation of complex audio data, introducing new capabilities that were previously unavailable. AFMs have the potential not only to revolutionize signal processing (SP) research and practice, but also to transform SP education by enabling innovative teaching methodologies and interactive learning experiences.

Previous efforts in computing and engineering education have leveraged audio-based multimedia approaches to enhance learning and foster intuition for abstract mathematical concepts. For example, \textit{DSP First} by McClellan, Schafer, and Yoder \cite{mcclellan2015dsp} integrated hands-on MATLAB labs with interactive exercises to engage students in signal processing and electrical engineering. Similarly, the \textit{Media Computation} approach by Guzdial and Ericson \cite{guzdial2020python} used audio manipulation to make programming more accessible to novice learners. These approaches have laid a strong foundation for multimedia-based education by providing interactive and context-driven learning experiences. The emergence of AFMs presents an opportunity to extend SP and engineering education more broadly, by enabling new forms of interactive learning.
%The emergence of AFMs presents an opportunity to build on these successes and introduce new dimensions to SP education (and engineering education more broadly). Unlike traditional multimedia methods that rely on pre-designed content and static tools, AFMs enable real-time interaction with signals through advanced capabilities such as speech-to-text, multilingual audio processing, and emotion recognition. These advantages allow students to interact dynamically with concepts, fostering deeper conceptual understanding through experiential and active learning. For example, AFMs can enable students to visualize and manipulate signals in real-time without requiring extensive programming expertise, thus lowering barriers to entry and enhancing accessibility.

Efforts to integrate GenAI into engineering education have already demonstrated its potential to reshape teaching and learning. For example, Qadir \cite{qadir2023engineering} examined the promise and pitfalls of GenAI in education, highlighting its capability to personalize learning and automate feedback. Similarly, Johri et al. \cite{johri2023generative} discussed the implications of generative AI for engineering education, emphasizing its ability to enhance engagement and foster creativity. These works underline the potential of GenAI to revolutionize engineering education broadly, providing a strong motivation to explore its specific applications in SP education.

We explore two key themes in this paper: (1) positioning AFMs as a transformative technology for signal processing (SP) by integrating innovations from GenAI and SP, and (2) examining their potential to revolutionize SP education as a model for advancing engineering education. By merging the technological potential of AFMs with lessons from prior educational initiatives, this paper presents a roadmap for driving progress in both SP practice and education through GenAI.

% In this regard, our paper makes the following three key contributions. First, it explores the advanced capabilities of AFMs and their potential to overcome persistent challenges in traditional SP methods. Second, it highlights how AFMs can transform SP education by building on multimedia-based approaches to create more engaging, interactive, and accessible learning experiences. As part of this, a case study demonstrates the feasibility of using AFMs for interactive auditory demonstrations in a DSP course, showcasing their impact on comprehension and engagement. Finally, the paper examines critical challenges such as ethical considerations, explainability, and customization, offering practical recommendations for integrating AFMs into SP curricula and paving the way for broader adoption of GenAI in engineering education.

In this regard, we make the following key contributions:
\begin{itemize}
    %\item We explore the advanced capabilities of AFMs and their potential to address persistent challenges in traditional SP methods.
    \item We demonstrate how AFMs can transform SP education by enhancing multimedia-based learning approaches, making them more engaging, interactive, and accessible. To support this, we present a case study showcasing the use of AFMs for interactive auditory demonstrations in a DSP course, highlighting their impact on comprehension and engagement.
    \item We examine critical challenges such as ethical considerations, explainability, and customization, offering practical recommendations for integrating AFMs into SP curricula and paving the way for the broader adoption of GenAI in engineering education.
\end{itemize}

\section{State-of-the-Art in Audio Foundation Models}
The advent of AFMs mark a significant advancement in audio signal processing. They not only surpass traditional models, relying on handcrafted features and shallow approaches, but also outperform conventional sequence models through large-scale pre-training and transformer-based architectures. AFMs like AudioLM \cite{borsos2023audiolm} and SpeechGPT \cite{zhang2023speechgpt} demonstrate strong cross-task generalization due to pre-training on diverse datasets. They excel in tasks such as automatic speech recognition (ASR), text-to-speech (TTS), audio synthesis, machine translation, speaker identification, emotion recognition, audio event detection, voice conversion, and music generation. Their capabilities also extend to real-time transcription, speech feedback, and interactive audio, enabling applications in accessibility, content creation, and personalized learning.

% The advent of AFMs marks a significant advancement in the field of audio signal processing. These models move beyond traditional approaches that relied on task-specific, handcrafted feature extraction and shallow models, by leveraging large-scale pre-training and transformer-based architectures. AFMs like AudioLM \cite{borsos2023audiolm} and SpeechGPT \cite{zhang2023speechgpt} exhibit strong cross-task generalization due to pre-training on massive, diverse datasets, making them well-suited for a variety of tasks such as automatic speech recognition (ASR), text-to-speech (TTS), audio synthesis, and machine translation. 

% \hl{(audio enhancement? audio source separation? https://dcase.community/challenge2024/task-language-queried-audio-source-separation)}

\begin{figure*}[!t]
\centering
\resizebox{0.82\textwidth}{!}{ % scale factor here to fit in the column and avoid overlaps
\begin{tikzpicture}[
    mindmap,
    grow cyclic,
    text width=2.8cm, % width slightly for each node
    concept color=cyan!30,
    level 1/.append style={level distance=6.5cm, sibling angle=60}, % level distance to create more space
    level 2/.append style={level distance=6.5cm, text width=2.5cm, font=\bfseries}, % inner text smaller and adjust distances
    every node/.style={concept, align=center, font=\bfseries}, % Small font size for better fitting
    root concept/.append style={
        concept color=teal!20, 
        font=\large\bfseries
    }
]

% Central node
\node[root concept] {Signal Processing\\ Techniques for AFMs}
    % Level 1 nodes (categories)
    child[concept color=lime!30, grow=80] { % sibling angle slightly to space nodes better
        node {Self-Supervised Learning}
        % Level 2 nodes (subcategories for Self-Supervised Learning)
        child[concept color=lime!15, grow=70] { node {Masked Signal Prediction\\(Wav2Vec 2.0)} }
        child[concept color=lime!15, grow=110] { node {Denoising\\(Whisper)} }
        child[concept color=lime!15, grow=150] { node {Contrastive Learning\\(HuBERT)} }
    }
    child[concept color=yellow!40, grow=20] { %  angles for spacing
        node {Autoencoders}
        % Level 2 nodes (subcategories for Autoencoders)
        child[concept color=yellow!20, grow=10] { node {Denoising Autoencoders\\(DeepSpeech)} }
        child[concept color=yellow!20, grow=45] { node {Variational Autoencoders\\(AudioLM)} }
        child[concept color=yellow!20, grow=80] { node {Sparse Autoencoders\\(General Use)} }
    }
    child[concept color=red!40, grow=-30] { %  sibling angle to prevent overlap
        node {Feature Extraction}
        % Level 2 nodes (subcategories for Feature Extraction)
        child[concept color=red!20, grow=20] { node {Fourier Transform\\(Whisper)} }
        child[concept color=red!20, grow=-15] { node {Log-Mel Spectrograms\\(Whisper, VGGish)} }
        child[concept color=red!20, grow=-40] { node {Wavelet Transform\\(AudioLM)} }
        child[concept color=red!20, grow=-75] { node {Vector Quantization (VQ)\\(Wav2Vec 2.0)} }
        child[concept color=red!20, grow=-115] { node {Residual Vector Quantization (RVQ)\\(Wav2Vec 2.0)} }
    }
    child[concept color=violet!30, grow=-105] { %  sibling angle to prevent overlap
        node {Multimodal Processing}
        % Level 2 nodes (subcategories for Multimodal Processing)
        child[concept color=violet!15, grow=-130] { node {Audio-Text Integration\\(Whisper)} }
        child[concept color=violet!15, grow=-165] { node {Audio-Visual Integration\\(AudioLM)} }
        child[concept color=violet!15, grow=-200] { node {Cross-Modal Learning\\(AudioLM)} }
    }
    child[concept color=orange!40, grow=135] { %  sibling angle to create space
        node {Noise Reduction}
        % Level 2 nodes (subcategories for Noise Reduction)
        child[concept color=orange!20, grow=150] { node {Adaptive Filtering\\(HuBERT)} }
        child[concept color=orange!20, grow=185] { node {Spectral Subtraction\\(DeepSpeech)} }
        child[concept color=orange!20, grow=220] { node {Blind Source Separation\\(DeepSpeech)} }
    };

\end{tikzpicture}
}
\caption{Taxonomy of Signal Processing Techniques for Audio Foundation Models (AFMs).} %The figure displays key signal processing techniques for AFMs, categorized into areas like Self-Supervised Learning, Autoencoders, and Feature Extraction, with models such as Whisper and Wav2Vec 2.0 linked to each technique}
\label{fig:afm_signal_processing}
\end{figure*}
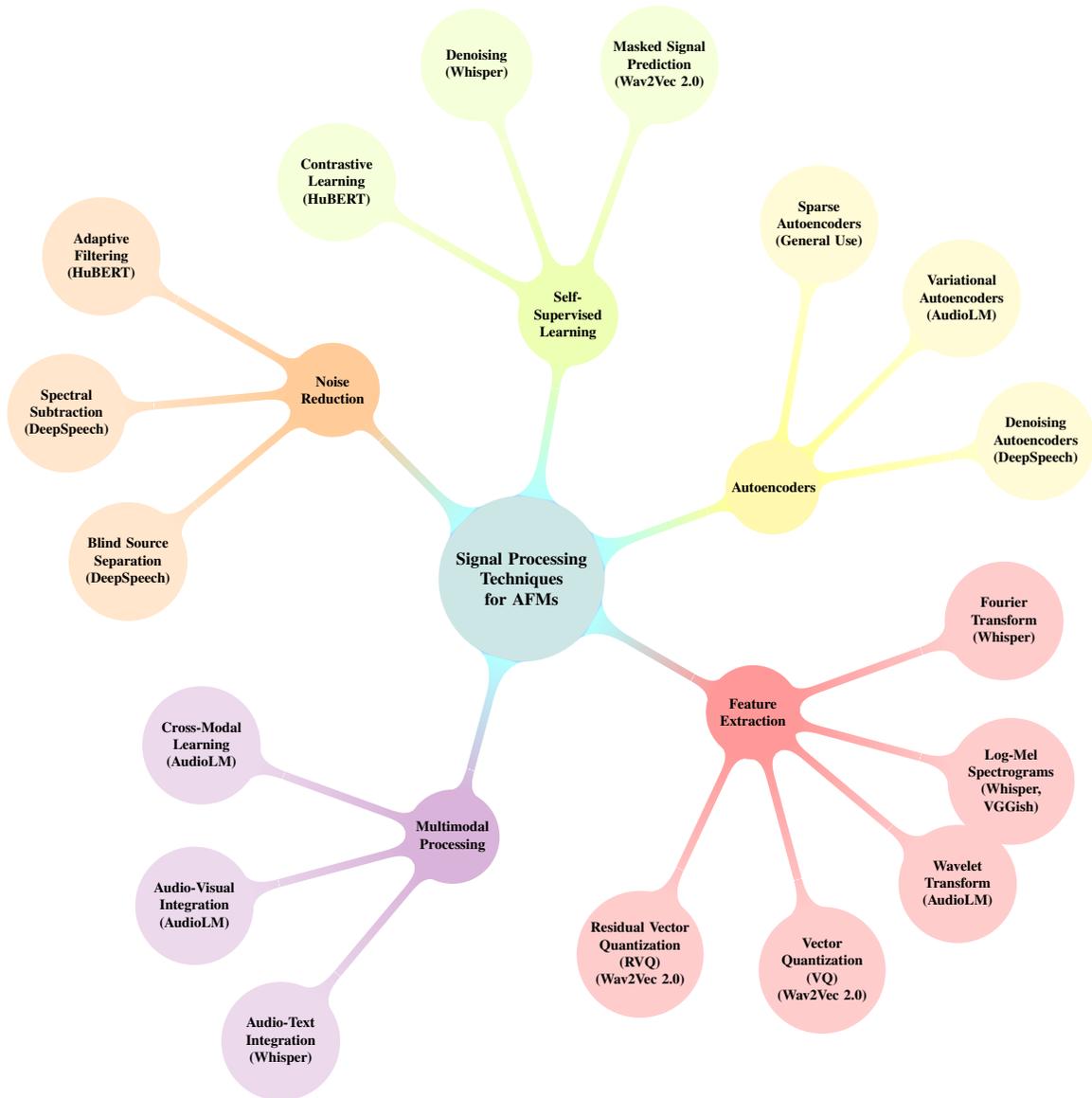

\subsection{Key Characteristics of AFMs}

\textit{What are Audio Foundation Models?} Audio Foundation Models (AFMs) are large-scale, transformer-based models pre-trained on extensive and diverse audio datasets. Unlike conventional models that rely on handcrafted features such as Mel-frequency cepstral coefficients (MFCCs), AFMs leverage deep learning architectures, particularly transformers \cite{vaswani2017attention}, to learn hierarchical and contextual representations directly from raw or minimally processed audio signals.
% AFMs are large-scale, transformer-based models pre-trained on extensive and diverse audio datasets. Unlike conventional models that rely on handcrafted features like Mel-frequency cepstral coefficients (MFCCs), AFMs use architectures such as transformers \cite{vaswani2017attention} to learn generalized audio representations. 
This enables AFMs to capture long-range dependencies in audio data, facilitating a deeper understanding of temporal and contextual relationships in speech and audio signals. To systematize the diverse signal-processing approaches underpinning AFMs, we conducted a focused review of foundational models (Wav2Vec 2.0, Whisper, HuBERT, AudioLM, DeepSpeech) and the associated technical literature. Figure \ref{fig:afm_signal_processing} synthesizes these insights into a taxonomy of core signal-processing techniques, grouping them into self-supervised learning, autoencoder-based methods, feature extraction pipelines, multimodal processing, and noise-reduction strategies. These categories are not exhaustive but are designed to capture the most recurrent and influential mechanisms that shape AFM performance across tasks.

% Figure \ref{fig:afm_signal_processing} further categorizes the essential signal processing techniques that can be used to optimize AFMs, including self-supervised learning, autoencoders, feature extraction, multimodal processing, and noise reduction, highlighting the interplay of AFMs with classical and advanced signal processing methods. 

\textit{Use of Self-Supervision and Pre-Training}: Self-supervised learning and pre-training are fundamental to AFMs, enabling them to generalize effectively across diverse tasks such as ASR, TTS, and music generation. 
% By being pre-trained on vast and heterogeneous datasets, AFMs learn robust audio representations that capture underlying patterns in the data.
By leveraging large and heterogeneous datasets during pre-training, AFMs learn robust audio representations that capture underlying patterns and contextual representations in the data. This significantly reduces the reliance on large labeled datasets, making AFMs especially valuable in scenarios where labeled data is scarce. Self-supervised learning allows AFMs to extract meaningful features from raw audio, leveraging unlabeled data to improve model performance. As a result, pre-trained AFMs can be fine-tuned for various downstream tasks with minimal task-specific data, shifting the focus from narrow, task-specific models to flexible, multi-purpose solutions \cite{latif2023sparks}.

\subsection{Multimodal Capabilities and Signal Processing}

A defining feature of AFMs is their ability to integrate multiple modalities, such as audio and text, within a unified framework. 
%Unlike traditional models that required separate systems for each modality and relied on late fusion techniques, AFMs like AudioPaLM \cite{rubenstein2023audiopalm} and WavJourney \cite{liu2023wavjourney} process audio and text together, capturing intricate relationships and enabling deeper multimodal interactions. This seamless integration overcomes the limitations of conventional approaches, which often struggled to synchronize information across modalities, resulting in reduced performance. By employing shared representations, AFMs deliver robust, contextually-aware outputs, making them particularly effective for tasks like real-time translation of spoken content while preserving semantic context \cite{borsos2023audiolm}.
Unlike traditional models that required separate systems and late fusion techniques, AFMs like AudioPaLM \cite{rubenstein2023audiopalm} and WavJourney \cite{liu2023wavjourney} integrate audio and text processing, capturing intricate relationships and enabling deeper multimodal interactions. This seamless approach resolves synchronization issues, enhancing performance, and shared representations allow AFMs to deliver robust, contextually-aware outputs, making them ideal for tasks like real-time translation that preserves semantic context \cite{borsos2023audiolm}.
Similarly, incorporating vision and vision-language capabilities into AFMs can further enrich educational applications. Models like CLIP, GLIP, and SAM, which employ contrastive learning and prompting techniques, can be integrated with AFMs to analyze visual content in educational videos, recognize objects and scenes, and generate descriptions or captions, thereby creating more interactive and comprehensive learning experiences. 
%Multimodal agents, such as ChatVideo, further demonstrate the potential of AFMs to process and understand both audio and visual information from multi-channel videos. For instance, PandaGPT leverages ImageBind to incorporate audio and visual data, including depth, thermal, and IMU data, into LMMs. This integration enables a wide array of applications, such as generating different creative text formats based on various sensory inputs. Looking ahead, the development of AFMs with a unified interface for image and text inputs could unlock even more sophisticated interactions in educational settings. Students could provide both audio and visual inputs, such as spoken questions accompanied by images or diagrams, to receive more comprehensive and contextually relevant feedback \cite{CGV-110}.
Multimodal agents like ChatVideo illustrate AFMs' potential to process both audio and visual information from videos. PandaGPT, for example, uses ImageBind to incorporate data from audio, visuals, and other sensors, supporting creative text generation from diverse inputs. Looking ahead, AFMs with unified image and text interfaces could allow students to receive contextually relevant feedback through combined audio and visual inputs, like spoken questions with images \cite{CGV-110}.

% \subsection{Multimodal Capabilities and Signal Processing}

% A defining feature of AFMs is their capacity for multimodal integration, which allows them to handle and process multiple data types---particularly audio and text---within a unified framework. Unlike traditional approaches, which typically require separate models for each modality and late-stage fusion techniques to combine outputs, AFMs such as AudioPaLM \cite{rubenstein2023audiopalm} and WavJourney \cite{liu2023wavjourney} integrate audio and text processing seamlessly. This multimodal capability enables deeper interactions between these modalities, which is especially beneficial in applications requiring simultaneous processing of spoken language and written text, such as real-time translation, interactive tutoring, or educational feedback systems \cite{borsos2023audiolm}.

% Historically, multimodal signal processing struggled with synchronizing information across modalities, leading to performance limitations when using traditional models. These earlier approaches relied on separate systems for each modality and late fusion techniques to combine data streams. In contrast, AFMs employ shared representations that inherently capture the intricate relationships between different modalities, enabling robust and contextually aware outputs. For example, an AFM can perform real-time translation of spoken content while maintaining the original semantic context, providing a level of coherence and accuracy that conventional models find challenging to achieve.

\subsection{Recent Innovations in Audio Foundation Models}

\subsubsection{New AFM Developments and Tools}
Recent advances in AFMs, such as Google’s NotebookLM, have attracted considerable attention due to their enhanced natural language interaction capabilities and their ability to generate structured, conversational audio content derived from research documents. These developments reflect the ongoing evolution of AFMs, extending audio processing technologies toward more naturalistic and engaging user experiences.

\subsubsection{What's New and What Remains Constant}
Although AFMs introduce new capabilities like cross-task generalization and deeper multimodal integration, certain principles from traditional audio processing still apply:
\begin{itemize}
    \item \textit{Generalization Across Tasks}: AFMs can perform well across multiple tasks (e.g., ASR, translation, audio generation), unlike earlier models optimized for a single task.
    \item \textit{Cross-Modality Learning}: The ability to model relationships between audio and text is enhanced in AFMs, making them especially suitable for applications requiring integrated feedback.
    \item \textit{Signal Preprocessing}: AFMs can directly learn representations from raw audio in a data-driven manner. However, minimal preprocessing, such as spectrogram conversion, is still a popular choice in AFMs.
    % Preprocessing steps (e.g., spectrograms) are still used, though features are now learned in a data-driven manner.
    \item \textit{Task-Specific Fine-Tuning}: While AFMs generalize across tasks, fine-tuning is essential to achieve state-of-the-art performance in complex tasks such as speech emotion recognition \cite{shoukat2023breaking} and other domain-specific applications \cite{latif2023sparks}.
\end{itemize}

\section{Educational Applications of AFMs}

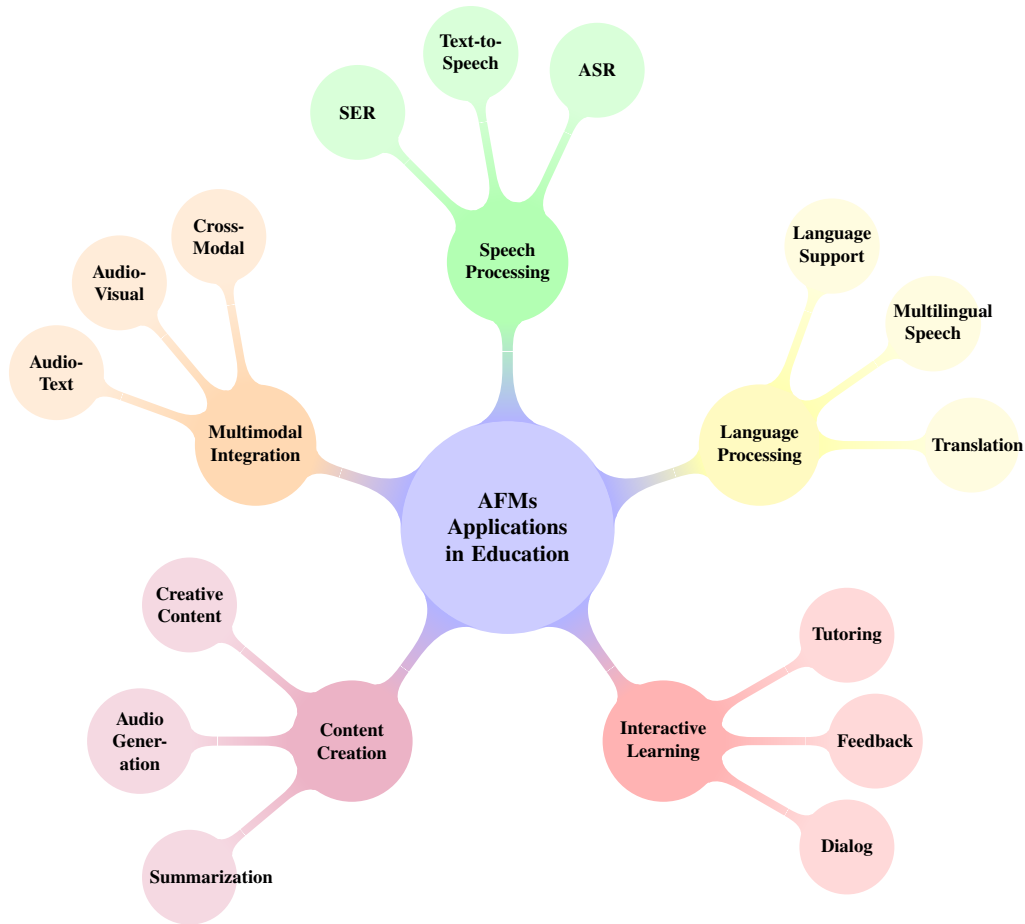
\begin{figure*}[!t]
\centering
\resizebox{0.80\textwidth}{!}{ % Adjust the scale factor here
\begin{tikzpicture}[
    mindmap,
    grow cyclic,
    text width=2.5cm,
    concept color=blue!30,
    level 1/.append style={level distance=5cm, sibling angle=100},
    level 2/.append style={level distance=4cm, sibling angle=100},
    every node/.style={concept, align=center, font=\bfseries},
    root concept/.append style={
        concept color=blue!20, 
        font=\large\bfseries
    }
]

% Central node
\node[root concept] {AFMs Applications\\ in Education}
    % Level 1 nodes (categories)
    child[concept color=green!30, grow=90] {
        node {Speech Processing}
        % Level 2 nodes (subcategories for Speech Processing) 
        child[concept color=green!15, grow=65] { node {ASR} }
        child[concept color=green!15, grow=100] { node {Text-to-Speech} }
        child[concept color=green!15, grow=135] { node {SER} }
    }
    child[concept color=yellow!30, grow=18] {
        node {Language Processing}
        % Level 2 nodes (subcategories for Language Processing)
        child[concept color=yellow!15, grow=0] { node {Translation} }
        child[concept color=yellow!15, grow=35] { node {Multilingual Speech} }
        child[concept color=yellow!15, grow=70] { node {Language Support} }
    }
    child[concept color=red!30, grow=-54] {
        node {Interactive Learning}
        % Level 2 nodes (subcategories for Interactive Learning)
        child[concept color=red!15, grow=-30] { node {Dialog} }
        child[concept color=red!15, grow=0] { node {Feedback} }
        child[concept color=red!15, grow=30] { node {Tutoring} }
    }
    child[concept color=purple!30, grow=-126] {
        node {Content Creation}
        % Level 2 nodes (subcategories for Content Creation)
        child[concept color=purple!15, grow=-140] { node {Summarization} }
        child[concept color=purple!15, grow=-180] { node {Audio Generation} }
        child[concept color=purple!15, grow=-220] { node {Creative Content} }
    }
    child[concept color=orange!30, grow=162] {
        node {Multimodal Integration}
        % Level 2 nodes (subcategories for Multimodal Integration)
        child[concept color=orange!15, grow=-200] { node {Audio-Text} }
        child[concept color=orange!15, grow=-230] { node {Audio-Visual} }
        child[concept color=orange!15, grow=-260] { node {Cross-Modal} }
    };

\end{tikzpicture}
}
\caption{Taxonomy of AFM Applications in Education}
\label{fig:afm_radial}
\end{figure*}

AFMs enhance both in-class and out-of-class learning through personalized instruction, automated assessment, and improved accessibility. Table \ref{table:afm_applications} and Figure \ref{fig:afm_radial} present a representative taxonomy of AFM functionalities and applications across key educational domains, illustrating how AFMs and signal processing techniques synergistically support more effective and engaging learning experiences. The following sections examine key applications and discuss how AFMs address diverse pedagogical challenges to support more effective and engaging learning experiences.

%AFMs greatly enhance both in-class and out-of-class learning by enabling personalized learning, automated assessments, and improved accessibility. Table \ref{table:afm_applications} outlines key AFM functionalities in education, such as real-time transcription, interactive dialogue, and automated language assessments. To complement this, Figure \ref{fig:afm_radial} provides a taxonomy of AFM applications, categorizing them into five areas: speech processing, language support, interactive learning, content creation, and multimodal integration, each with specific tasks that illustrate AFMs' versatility in transforming educational practices. This offers a comprehensive view of how AFM applications and SP techniques work synergistically to improve educational outcomes.This taxonomy is not intended to be exhaustive; rather, it provides a coherent, collectively representative view of how AFMs are being deployed across educational contexts. Next, we delve into the key applications and explore how AFMs address diverse pedagogical challenges, enabling more effective and engaging educational experiences.

\subsection{Personalized Education}

AFMs can enhance personalized learning by supporting educators in guiding critical thinking and socio-emotional development while preserving the teacher-student relationship. AI should provide real-time feedback and assistance but leave complex decision-making to human instructors to maintain meaningful interactions. AFMs also empower students with self-directed learning features, such as real-time transcription and sentiment analysis, fostering an engaging educational environment. 

\subsubsection{Inclusive Education and Accessibility}
%In particular, 
AFMs offer opportunities to bridge the educational divide by improving accessibility for students with disabilities or those in underdeveloped regions. AFMs can generate audio content in multiple languages, offer real-time speech-to-text transcriptions, and provide audio summaries for students with visual impairments.

\begin{table}[t]
\scriptsize
\centering
\caption{Applications of AFMs in Education}
\label{table:afm_applications}

\renewcommand{\arraystretch}{1.05}
\begin{tabularx}{\columnwidth}{|p{2.1cm}|X|p{2.1cm}|}
\hline
\textbf{Functionality} & \textbf{Educational Use Cases (Summary)} & \textbf{AFM Examples} \\
\hline

\textbf{Speech-to-Text \& Summarization} 
& Lecture transcription; summarization; note-taking. 
& Whisper; AudioLM \\
\hline

\textbf{Multilingual Speech Processing} 
& Real-time translation; multilingual TTS; support for non-native speakers.
& AudioPaLM; Voicebox \\
\hline

\textbf{Interactive Feedback \& Assessment} 
& Pronunciation correction; fluency evaluation; automated scoring. 
& SpeechGPT; Wav2Vec 2.0 \\
\hline

\textbf{Emotion Recognition \& Adaptation} 
& Emotion-aware feedback; adaptive instruction; emotional support. 
& HuBERT; Wav2Vec 2.0 \\
\hline

\textbf{Text-to-Speech \& Audio Generation} 
& Reading support; realistic simulations; creative audio tasks. 
& Tacotron 2; Jukebox; Audiobox \\
\hline

\textbf{Multimodal \& Hybrid Learning} 
& Multimodal lectures; synchronized materials; complex concepts. 
& AudioLM; Multimodal GPT \\
\hline

\end{tabularx}
\end{table}

Moreover, AFMs can generate content tailored to various educational levels, thus supporting equitable access to learning materials and reducing the digital divide.

\subsubsection{Intelligent Tutoring Systems (ITS)}
%One of the most promising applications of AFMs in education is in ITS. 
AFMs can power adaptive audio tutoring, where the system listens to students' verbal responses, processes their performance, and offers tailored instructional feedback. For instance, an ITS could use an AFM to provide real-time pronunciation correction during language learning or help students improve their mathematical problem-solving skills by guiding them through step-by-step explanations.

\subsubsection{Automated Feedback and Assessment}
AFMs can evaluate and provide feedback on spoken presentations, language fluency, and oral exams. This includes assessing pronunciation, tone, and clarity in language learning applications, making it possible to provide immediate and accurate feedback on students' performance without the need for constant human supervision \cite{latif2023sparks}. Additionally, AFMs can be employed to automatically transcribe and evaluate the quality of class discussions,  streamlining the feedback process.

\subsubsection{Engagement Through Gamification and Interaction}
AFMs can be integrated into educational games that use audio cues, interactive conversations, or storytelling to engage students. This approach can help improve attention, creativity, and problem-solving skills in data-driven learning environments. AFMs could, for example, power language-learning games where students converse with AI-powered characters that offer feedback and encourage language practice.

\subsection{Versatile Learning Inside and Outside the Classroom}

%AFMs play a pivotal role in enriching both in-class and out-of-class learning experiences. In the classroom, they enhance real-time interactions through applications like live transcription, automated audio feedback, and personalized tutoring. For instance, AFMs can provide immediate feedback on student presentations, saving time for educators while ensuring students receive meaningful, timely input \cite{ahmad2023education}. Additionally, high-quality speech-to-text capabilities improve accessibility for students with hearing impairments, making learning more inclusive. Outside the classroom, AFMs support personalized learning by offering tailored audio summaries, podcasts, and revision tools. Intelligent virtual tutors powered by AFMs provide adaptive learning experiences, engaging students in real time and accommodating different learning speeds to maintain student motivation and progress. Another valuable application of AFMs is in real time language translation during international collaboration or cultural studies. AFMs could provide students and educators with real-time translation of classroom discussions or lectures, breaking language barriers and enabling cross-cultural learning opportunities.

AFMs significantly enhance both in-class and out-of-class learning experiences. In the classroom, they improve real-time interactions through applications like live transcription, automated audio feedback, and personalized tutoring. For example, AFMs can give immediate feedback on student presentations, saving educators time while ensuring timely input for students \cite{ahmad2023education}. Additionally, high-quality speech-to-text capabilities enhance accessibility for students with hearing impairments. Outside the classroom, AFMs facilitate personalized learning by providing tailored audio summaries, podcasts, and revision tools. Intelligent virtual tutors powered by AFMs offer adaptive learning experiences, engaging students in real time and accommodating different learning speeds. AFMs also enable real-time language translation during international collaboration, helping break language barriers and promoting cross-cultural learning opportunities.
%\subsection{Supporting Cross-Cultural Learning}

\subsection{AFMs for Supporting Teachers}
Beyond student-focused applications, AFMs can assist teachers in managing their workload. For instance, AFMs can help grade oral exams, automate lecture transcriptions, and generate personalized feedback for students' spoken assignments. By offloading these time-consuming tasks, educators can focus more on direct interactions with students, fostering a more engaging and supportive learning environment.

\section{Transforming SP Education with AFMs: A Vision for \textbf{SPEduAFM}}

Inspired by the revolutionary impact of \textit{DSP First} \cite{mcclellan2015dsp} on signal processing education and the Python-based framework \textit{Think DSP} \cite{downey2016think}, specifically designed for DSP education, we envision the development of a specialized AFM tailored specifically for DSP education: \textit{SPEduAFM}. Both \textit{DSP First} and \textit{Think DSP} have demonstrated the pedagogical effectiveness of interactive and multimedia-based approaches in teaching complex signal processing concepts, setting the stage for the next evolution in DSP education.

%While similar approaches such as MATLAB-based and Python-based frameworks have been instrumental in DSP education, SPEduAFM represents a paradigm shift by utilizing state-of-the-art AFMs like Whisper (OpenAI), Wav2Vec 2.0 (Meta), and AudioLM (Google). This vision outlines both the potential applications of SPEduAFM and a phased approach to its development, highlighting the significant impact it can have on modern pedagogy.

While similar approaches such as MATLAB-based and Python-based frameworks have been instrumental in DSP education, SPEduAFM represents a paradigm shift by envisioning a specialized AFM tailored specifically for signal processing education. Instead of piecing together various existing AFMs like Whisper (OpenAI), Wav2Vec 2.0 (Meta), and AudioLM (Google), SPEduAFM could be designed from the ground up to address the unique needs of DSP education. This vision outlines both the potential applications of SPEduAFM and a phased approach to its development, highlighting the significant impact it can have on modern pedagogy.

Table \ref{tab:speeduafm_comparison} presents a comparative analysis of traditional MATLAB/Python-based SP labs and the envisioned SPEduAFM framework, highlighting how the latter leverages multimodal capabilities—audio and text commands—along with real-time interactivity to enrich learning. By integrating advanced AI tools and enabling hands-on experimentation, SPEduAFM not only addresses the limitations of conventional approaches but also unlocks new pedagogical possibilities previously unattainable.

\subsection{Vision for SPEduAFM}
\textit{SPEduAFM} is conceptualized as a comprehensive platform that integrates real-time interactivity, advanced pedagogical tools, and cutting-edge machine learning techniques. The platform aims to enable the following:

\begin{itemize}
    \item \textit{Dynamic Real-Time Interactions}: Students can manipulate audio signals in real time, applying concepts like Fourier analysis, filtering, and noise suppression while observing immediate effects.
    \item \textit{Lower Entry Barriers}: Unlike traditional tools requiring extensive programming expertise, SPEduAFM will provide an intuitive interface for seamless experimentation.
    \item \textit{Generative DSP Tools}: By incorporating AFMs, students can synthesize and analyze audio, exploring concepts like speech generation, emotion recognition, and multimodal signal integration.
\end{itemize}

\subsection{Integrating SPEduAFM into SP Lab Education}
Traditional SP education has evolved from mathematical formulations to hands-on MATLAB and Python-based implementations. However, the emergence of GenAI and AFMs presents an opportunity to further enhance learning by enabling real-time interaction, automation, and multimodal integration. In this context, the proposed SPEduAFM, aims to transition DSP education from conventional computational methods to AI-powered, interactive, and personalized learning experiences. Table \ref{tab:speeduafm_comparison} compares traditional DSP lab approaches with the AI-driven methodologies enabled by SPEduAFM.

\begin{table*}[htp]
\caption{Comparison of Traditional DSP Labs and AFM-Based SPEduAFM Approach}
\label{tab:speeduafm_comparison}
\scriptsize
\resizebox{\textwidth}{!}{%
\begin{tabular}{|c|p{2.3cm}|p{5.2cm}|p{6.8cm}|p{4.2cm}|p{1.8cm}|}
\hline
\textbf{Sr. No} & \textbf{Lab Experiment} & \textbf{Traditional Approach \newline (MATLAB/Python)} & \textbf{AFMs Approach} & \textbf{Pedagogical Advancements with AFM} & \textbf{AFMs Utilized} \\
\hline
1 & Signal Generation \& Visualization & \textbf{MATLAB}: Generate sine/cosine waves using \texttt{sin()} and \texttt{cos()} and plot with \texttt{plot()}. \textbf{Python}: Use \texttt{numpy.sin()}, \texttt{numpy.cos()}, and \texttt{matplotlib.pyplot.plot()}. & \textbf{SPEduAFM}: Generate sine/cosine waves using natural language commands through: 
\textit{1) Audio Command}: Speak, e.g., ``Generate a 440 Hz sine wave.'' 
\textit{2) Text Command}: Input text referencing an audio signal, e.g., ``Generate waveform for audio\_file.wav.'' Results are visualized in real time via a dashboard. & Enables multimodal input (audio and text), allowing hands-free signal generation and immediate feedback using interactive spectrograms. & AudioLM, Whisper \\
\hline
2 & Fourier Transform \& Spectral Analysis & \textbf{MATLAB}: Compute FFT using \texttt{fft()} and visualize with \texttt{stem()}. \textbf{Python}: Compute FFT using \texttt{numpy.fft.fft()} and plot magnitude spectra using \texttt{matplotlib}. & \textbf{SPEduAFM}: Perform FFT with commands given as: 
\textit{1) Audio Command}: Speak, e.g., ``Compute the FFT of a 50 Hz sine wave.'' 
\textit{2) Text Command}: Reference a recorded signal, e.g., ``Compute the FFT of audio\_file.wav and plot the spectrum.'' & Provides real-time analysis of spectral content and interactive exploration of frequency-domain transformations. & Whisper, Wav2Vec 2.0 \\
\hline
3 & Filtering \& Noise Removal & \textbf{MATLAB}: Design FIR/IIR filters using \texttt{designfilt()} and apply using \texttt{filter()}. \textbf{Python}: Use \texttt{scipy.signal.firwin()} and \texttt{iirfilter()}. & \textbf{SPEduAFM}: Apply FIR/IIR filters interactively via: 
\textit{1) Audio Command}: Speak, e.g., ``Apply a low-pass FIR filter with a 100 Hz cutoff.'' 
\textit{2) Text Command}: Reference an audio signal, e.g., ``Filter audio\_file.wav with a low-pass FIR filter.'' & Enhances comprehension by enabling real-time auditory output and parameter adjustments through voice or text referencing audio. & Whisper, SpeechGPT, DeepSpeech \\
\hline
4 & Speech Processing \& Feature Extraction & \textbf{MATLAB}: Extract MFCCs using \texttt{mfcc()}. \textbf{Python}: Use \texttt{librosa.feature.mfcc()} for extraction. & \textbf{SPEduAFM}: Extract MFCCs and spectrograms through: 
\textit{1) Audio Command}: Speak, e.g., ``Extract MFCCs from my recorded voice.'' 
\textit{2) Text Command}: Reference audio data, e.g., ``Compute MFCCs for audio\_file.wav and display features.'' & Strengthens understanding of speech features by integrating audio-visual learning and dynamic feature extraction. & HuBERT, Whisper \\
\hline
5 & Adaptive Filtering (LMS, RLS) & \textbf{MATLAB}: Implement LMS using \texttt{dsp.LMSFilter()}. \textbf{Python}: Use \texttt{scipy.signal.lfilter()}. & \textbf{SPEduAFM}: Apply adaptive filtering via: 
\textit{1) Audio Command}: Speak, e.g., ``Apply LMS filter to this live audio and reduce noise.'' 
\textit{2) Text Command}: Reference recorded data, e.g., ``Run LMS adaptive filter on audio\_file.wav.'' & Enables practical learning by exploring filter tuning through live audio inputs or pre-recorded signals. & HuBERT, DeepSpeech \\
\hline
6 & Convolution \& \newline Correlation & \textbf{MATLAB}: Perform convolution using \texttt{conv()} and correlation using \texttt{xcorr()}. \textbf{Python}: Use \texttt{numpy.convolve()} and \texttt{numpy.correlate()}. & \textbf{SPEduAFM}: Execute convolution and correlation via: 
\textit{1) Audio Command}: Speak, e.g., ``Convolve my voice with a high-pass filter.'' 
\textit{2) Text Command}: Reference recorded signals, e.g., ``Convolve audio\_file.wav and plot results.'' & Promotes deeper understanding through real-time experimentation and visualization of convolution effects on speech and signals. & Whisper, AudioLM \\
\hline
7 & Wavelet Transform \& Time-Frequency Analysis & \textbf{MATLAB}: Perform wavelet decomposition using \texttt{cwt()} and \texttt{dwt()}. \textbf{Python}: Use \texttt{pywavelets.wavedec()} for wavelet decomposition. & \textbf{SPEduAFM}: Perform wavelet analysis on audio signals via: 
\textit{1) Audio Command}: Speak, e.g., ``Apply Continuous Wavelet Transform (CWT) to this signal.'' 
\textit{2) Text Command}: Input, e.g., ``Analyze audio\_file.wav using wavelet decomposition.'' & Provides real-time interaction with wavelet decomposition, aiding comprehension of time-frequency trade-offs. & AudioLM, SpeechGPT \\
\hline
8 & Real-Time Speech Enhancement & \textbf{MATLAB}: Implement noise suppression using Wiener filters or spectral subtraction. \textbf{Python}: Use \texttt{scipy.signal.wiener()} for noise reduction. & \textbf{SPEduAFM}: Enhance speech quality dynamically via: 
\textit{1) Audio Command}: Speak, e.g., ``Suppress background noise in this audio.'' 
\textit{2) Text Command}: Input, e.g., ``Denoise audio\_file.wav and improve clarity.'' & Demonstrates real-world noise suppression techniques, allowing comparison of enhancement methods in real time. & HuBERT, DeepSpeech \\
\hline
9 & Multimodal DSP Applications & \textbf{MATLAB}: Process audio and images separately using \texttt{audioread()} and \texttt{imshow()}. \textbf{Python}: Use \texttt{opencv()} and \texttt{librosa()} for separate processing. & \textbf{SPEduAFM}: Perform multimodal processing via: 
\textit{1) Audio Command}: Speak, e.g., ``Analyze audio and match with video frames.'' 
\textit{2) Text Command}: Input, e.g., ``Describe facial expressions in video\_file.mp4 and map them to speaker emotion in audio\_file.wav.'' & Expands DSP education beyond audio-only applications, reinforcing multimodal AI understanding and cross-modal learning. & Whisper, AudioLM, Multimodal GPT \\
\hline
\end{tabular}%
}
\end{table*}

\subsection{Applications Enabled by SPEduAFM}
One of the envisioned applications of SPEduAFM is an \textit{Interactive Auditory Demonstration Platform} for teaching DSP concepts. This platform would allow students to:
\begin{itemize}
    \item Experiment with adaptive filters on live audio streams to understand noise cancellation techniques.
    \item Analyze speech signals for multilingual transcription and phoneme alignment, demonstrating practical applications of DSP.
    \item Visualize and interpret time-frequency transformations, such as wavelet analysis, to better understand signal decomposition.
    \item Investigate advanced concepts like multimodal integration, combining audio, text, and visual data for tasks like emotion recognition and transcription alignment.
\end{itemize}

These applications extend beyond static demonstrations to provide a rich, interactive learning experience that bridges theory and practice.

\subsection{Phased Development Approach}
To bring \textit{SPEduAFM} to fruition, we propose a phased approach:

\subsubsection{Phase 1: Fine-Tuning Existing AFMs}
In the initial phase, existing AFMs such as Whisper and AudioLM will be fine-tuned for educational tasks, focusing on DSP-specific datasets. This phase will also involve developing basic interactive tools for real-time signal processing and spectrogram visualization.

\subsubsection{Phase 2: Platform Development}
A web-based platform will be developed with an intuitive interface, multilingual support, and capabilities for advanced DSP tasks like adaptive filtering and wavelet transforms. FastAPI integration will ensure real-time processing, while backend optimizations minimize latency.

\subsubsection{Phase 3: Bespoke SPEduAFM Model Creation}
With additional resources and datasets, a bespoke \textit{SPEduAFM} can be developed to address the specific needs of DSP education. This model will be tailored to provide generative capabilities, multimodal integration, and advanced pedagogical tools.

\subsection{Interactive Demonstrations for Experiential Learning}
By integrating advanced AFMs, \textit{SPEduAFM} can enable students to engage in hands-on learning activities, including:
\begin{itemize}
    \item Applying adaptive filters to live audio signals to explore noise cancellation techniques and understand practical DSP applications.
    \item Synthesizing speech or environmental sounds and analyzing their spectral properties to grasp core DSP principles like Fourier analysis.
    \item Manipulating audio streams to simulate real-world DSP challenges such as noise suppression or dynamic equalization.
    \item Visualizing audio decomposition through advanced tools like wavelet transforms, offering a deeper understanding of time-frequency analysis.
    \item Comparing traditional DSP algorithms with AFM-based approaches to critically evaluate their trade-offs and applications.
\end{itemize}

\subsection{Idealized Vision for SPEduAFM}
The ultimate goal is to create an inclusive, accessible, and scalable platform that redefines DSP education. By aligning with flipped and blended learning methodologies \cite{van2013flipping, bajwa2017flipping}, \textit{SPEduAFM} aims to:
\begin{itemize}
    \item Enhance engagement and retention by offering interactive, real-time feedback.
    \item Enable students to explore complex DSP concepts without the steep learning curve of traditional programming-based tools.
    \item Foster critical thinking by allowing comparisons between classical DSP methods and AFM-driven approaches.
    \item Provide a modular framework for educators to adapt the platform to their specific curriculum needs.
\end{itemize}

\textit{SPEduAFM} represents a transformative vision for DSP education, building on the successes of \textit{DSP First} and leveraging the capabilities of generative AI. While initial efforts can rely on fine-tuning existing AFMs, the development of a bespoke \textit{SPEduAFM} can unlock unprecedented opportunities for experiential learning, bridging theoretical knowledge with real-world applications. By providing a dynamic and inclusive platform, \textit{SPEduAFM} has the potential to inspire a new generation of engineers and researchers in signal processing.

\section{Open Research Issues and Future Directions}

AFMs and multimodal AI hold transformative potential for education, but their deployment must overcome key challenges to ensure ethical and effective use \cite{qadir2023engineering}, which we discuss next. 

\subsection{Innovations in GenAI and Signal Processing}

\subsubsection{Advanced Self-Supervised Learning Techniques}

Self-supervised learning approaches, such as masked signal modeling and contrastive learning, can be enhanced by incorporating more sophisticated data augmentation and denoising techniques tailored specifically for audio domains. This could involve creating new forms of data distortion that simulate real-world audio challenges, thereby improving the robustness and generalization of AFMs.

\subsubsection{Hybrid Architectures}

In SP education, mastering fundamental concepts such as those outlined in the Signals and Systems Concept Inventory (SSCI) \cite{wage2005signals} is critical. These concepts, grouped into categories like background mathematical concepts, linearity and time invariance, convolution, transform representations, filtering, and sampling, form the backbone of SP curricula. By leveraging hybrid architectures that combine traditional signal processing techniques (e.g., wavelet transforms, spectral analysis) with advanced deep learning and transformer models, AFMs can be optimized to support SP education. Neural architecture search and automated feature engineering methods can further enhance these hybrid models, balancing performance with computational efficiency. This approach ensures that AFMs are well-suited for resource-constrained environments, enabling customized educational experiences for SP students.

\subsubsection{Information Retrieval for Subject Matter Knowledge}

Building effective GenAI systems for SP education requires a robust foundation of subject matter knowledge. Tools like the SSCI \cite{wage2005signals} provide a structured framework of essential SP concepts, but further enrichment is possible by incorporating other domain-specific resources, such as course syllabi, textbooks, lecture slides, and even instructor annotations. Using advanced information retrieval techniques, these resources can be integrated into the GenAI system to create a comprehensive knowledge base tailored to a specific SP course.

For instance, Retrieval-Based Generation (RAG) techniques can dynamically fetch relevant information from this knowledge base to contextualize and personalize the output generated by AFMs. A pipeline can be designed to combine AFMs' audio processing capabilities with text-based information retrieval, creating multimodal GenAI systems capable of delivering customized learning content. These systems can provide detailed explanations, solve problems, or generate practice exercises aligned with the specific learning objectives of a course, significantly enhancing the learning experience for students.

\subsubsection{Multimodal Learning and Audio Fusion}

Integrating multimodal learning approaches with advanced audio fusion techniques can enhance AFMs' ability to learn from diverse data modalities, such as audio, text, and video. For example, a multimodal GenAI system could present audio-based explanations of SP concepts, such as sampling or convolution, augmented by visual aids like signal plots or step-by-step textual explanations. This fusion of modalities enables a richer, more intuitive learning experience, helping students bridge the gap between theoretical knowledge and its practical applications in SP.

\subsubsection{Real-Time Adaptation and Personalized Learning}

Adaptive learning algorithms can enable real-time fine-tuning of AFMs based on user feedback, audio inputs, or student performance metrics. For instance, a GenAI system powered by AFMs and enriched with subject matter knowledge could dynamically adjust content delivery based on a student's understanding of key SP topics, as assessed through tools like the SSCI. Such systems can offer tailored explanations or exercises, focusing on challenging areas like transform representations or filtering, to meet individual learning needs. Techniques for online learning and continuous model updates further ensure that these systems remain current with advances in SP pedagogy and technology. This adaptability fosters a personalized and engaging educational environment, empowering students to master SP concepts more effectively.

\subsection{Human-Centric Pedagogy and Designing Experiential GenAI Education}

Recent competency frameworks, such as those discussed by Cukurova et al. \cite{cukurova2024ai} and Miao and Shiohira \cite{UNESCO2024}, emphasize the importance of aligning AI tools with human-centric pedagogical approaches. These frameworks extend AI literacy discussions by addressing key competencies for teachers and students:

\begin{itemize}
    \item \textit{Teachers:} Educators must be equipped with skills to ethically integrate AI into pedagogy, emphasizing a human-centered mindset, understanding AI foundations, and continuous professional development.
    \item \textit{Students:} Learners should gain competencies in understanding AI’s technical foundations, ethical implications, and the design of inclusive AI systems to ensure responsible usage and innovation.
\end{itemize}

Building on these principles, the GUIDES framework \cite{QADIR2024GUI} offers a multifaceted approach to integrating GenAI tools into education. Particularly, the ``E'' in GUIDES stands for Experiential Engineering Education, advocating for hands-on, material-centric learning. Engineering education must leverage GenAI not only for theoretical exploration but also to enhance students' interaction with the physical world. This involves harmonizing the abstract computational capabilities of GenAI with real-world, material-centric experiences that are fundamental to engineering disciplines.

Authentic education, as highlighted by Qadir and Al-Fuqaha \cite{qadir2020student}, engages students in solving complex, open-ended problems that mirror real-world contexts, emphasizing realism, judgment, and active participation. By integrating AFM-based experiential education, abstract mathematical concepts and equations can be brought to life, motivating students and showcasing the practical relevance of theoretical knowledge. Wiggins and McTighe \cite{wiggins2005understanding} further stress that authentic tasks help students organize knowledge around ``big ideas,'' facilitating the transfer of understanding across contexts. For example, AFM-based systems can simulate real-world scenarios where students apply signal processing techniques to analyze noisy audio or design filters, enabling them to co-construct knowledge and master SP concepts meaningfully and enduringly.

By aligning with the experiential dimension, GenAI can:
\begin{itemize}
    \item Simulate real-world scenarios, such as using DSP for noise reduction or medical imaging.
    \item Provide interactive, multimodal explanations that connect abstract theories with practical applications.
    \item Empower students to test theoretical principles through virtual labs and simulations, bridging the gap between computation and physical experimentation.
\end{itemize}

Moreover, privacy-preserving and interpretable AI techniques are critical for maintaining trust in GenAI systems. Signal processing methods, such as attention visualization and interpretable embeddings, enhance explainability, while privacy-preserving strategies like differential privacy and federated learning protect sensitive data. These tools ensure that educational technology remains ethical, inclusive, and accessible, particularly in resource-constrained settings, as highlighted by UNESCO \cite{unesco2023genai}.

Ultimately, GenAI must serve as a tool to enrich human-centric pedagogy, fostering deep engagement with both theoretical and practical aspects of education. By integrating experiential and human-centered approaches, GenAI can empower students and educators alike, making education more inclusive, effective, and aligned with the demands of the modern world.

% \section{Conclusions}
% 

\section{Conclusions}
%Audio Foundation Models (AFMs), an emerging category of Generative AI (GenAI), promise to enhance education in general and signal processing education in particular. 
%This paper explores their versatility in performing diverse auditory tasks, including speech recognition, multilingual transcription, and real-time auditory feedback, offering novel pedagogical opportunities beyond traditional computational tools like MATLAB and Python. Unlike these conventional approaches, which primarily help students develop intuition through example-driven exploration, AFMs enable more interactive, immersive, and dynamic learning experiences. 

Audio Foundation Models (AFMs), an emerging form of Generative AI, are enhancing signal processing education. Their versatility in diverse auditory tasks offers novel pedagogical opportunities beyond traditional tools like MATLAB and Python, enabling more interactive, immersive, and dynamic learning experiences. SPEduAFM, introduced in this work, envisions a specialized AFM that integrates GenAI-driven innovations with core signal processing principles. By outlining its potential applications, we demonstrate how AFMs can bridge theoretical concepts and practical applications, making abstract SP topics more accessible and engaging. However, realizing this vision requires addressing key challenges, including ethical considerations, explainability, and adaptability to diverse educational contexts. By positioning SPEduAFM as a conceptual framework, this work aims to inspire further research and development in AI-driven educational tools, fostering innovative, inclusive, and human-centered learning experiences in signal processing and beyond.

\section{Acknowledgements}
The authors gratefully acknowledge support from Qatar University and partial support from  Grant no. MME04-0607-230060 from the Qatar Research, Development and Innovation (QRDI) Council, in collaboration with the Ministry of Municipality, Qatar. The findings herein reflect the work, and are solely the responsibility, of the authors. AI-based tools (ChatGPT and Grammarly) were utilized to assist with editing and enhancing the grammar of this paper.

\bibliographystyle{IEEEtran}
\bibliography{LargeAudioModels, AIinEducation}

\end{document}